# Technology For Information Engineering (TIE): A New Way of Storing, Retrieving and Analyzing Information

**Jerzy Lewak[1]**

**SpeedTrack, Inc., Nisus® Software Inc., & Department of Electrical and Computer Engineering, UCSD La Jolla CA 92093-0407**

## Abstract

The theoretical foundations of a new model and paradigm (called TIE) for data storage and access are introduced. Associations between data elements are stored in a single Matrix table, which is usually kept entirely in RAM for quick access. The model ties together a very intuitive "guided" GUI to the Matrix structure, allowing extremely easy complex searches through the data. Although it is an "Associative Model" in that it stores the data associations separately from the data itself, in contrast to other implementations of that model TIE guides the user to only the available information ensuring that every search is always fruitful. Very many diverse applications of the technology are reviewed.

# 1    Introduction

The technological revolution is inhibited by the Information explosion. Large amounts of information in electronic form are now available, but still very difficult to access. This paper introduces a new system for solving many problems associated with this difficulty.

A new technology initially called Guided Information Access or GIA and later dubbed "Technology for Information Engineering" or TIE, invented by the author and others in 1991 and patented [Lewak et al 1996] in very subtle ways changes current models of information, indexing, analysis and information access.

This subtle change has a profound influence on the speed and ease of access to every kind of

[1] Jerzy Lewak C/O Nisus Software Inc.

107 South Cedros Ave., Solana Beach, CA 92075

jerzy@nisus.com

information. It further influences greatly the theoretical analysis of information indexing, retrieval and organization. It has even broader applications such as pattern recognition, pattern matching in general and possibly even the mapping of the brain.

In general terms, though hard to describe, it can be thought of as a new kind of user interface to data, driven by novel kinds of software structures and procedures. It would never have been invented if thinking about the user interface were to be detached from the details of the software structures and implementation. It is very misleading to label it a new kind of search engine, though the temptation is there.

The TIE system is in fact similar to the "Associative Model" [Lazy Software 2000] in that it stores and uses Associations between Items of information. However, it is considerably simpler in its search syntax making it very much easier to search for data in TIE than in any present implementations of the said Associative Model.

A general way to view the TIE system is as a simple, almost grammarless language describing each Item of information in a database. The description is simply an association, without regard to order, of the descriptive words or phrases, called Categories, with the Item of information. The union set of all these Categories in a database forms the vocabulary of the language, specific to that database and is presented to the user to pick from in any order, organized in a structure that reflects the utility and the nature of the data.

What makes the system powerful is the rule, imposed by the implementation, that at all times the user can *fruitfully* choose any word or phrase listed, adding it to the current description. Here, *fruitfully* means that any choice made by the user will find some Items in the database.

Typically the user chooses one Category at a time, with either a mouse click or keyboard, and the system instantly responds by adjusting the vocabulary and removing the unavailable Categories. Thus the user is very naturally led to the available information, and only to that *available* information. In addition, the displayed vocabulary is always representative of the description of the available information, giving the user a "view" of the information - leading to useful queries that would otherwise never have occurred to the user to submit.

More generally, TIE allows the average user, completely transparently, to pose and instantly obtain responses to, complex boolean queries using the vocabulary of Categories.

It is important to recognize that Categories can be descriptive at many different levels of detail, as the context demands. For example, the individual words in a description of an Item can be Categories, or alternatively the whole description can be a Category. (Sometimes even both possibilities are used.) A more "atomic" level of detail is also possible, where, for example, the individual letters of the alphabet can be Categories. Likewise, the individual digits of a number can also be Categories. Both these are called "Alpha-Numeric Categories" or simply "Alpha Categories." Both these examples are currently being used in implementations in at least two ways: the Position Independent Alpha and the Position Dependent  Alpha Category groups. The Position Dependent Categories use the position in a string of letters or numbers as a Category attribute.

One technical way to view the system is to contrast its structures with those of the most prevalent current data-organization system. Current systems use hierarchical or tree structures. The TIE system uses the much more general graph structure. It is well known that [Hart 1992] our brains remember things through overlapping descriptive categories. Any implementation which uses this fact must lead to structures which are unrestricted by the hierarchy and so are more general then that. Graphs are structures of that type.

Current information retrieval methods, or searches, may be divided into two types:

    1   Structured searches

    2   Unstructured searches

Structured searches can be further divided into two types:

    i   Search through hierarchical directories or "folders"

    ii   Search by specific data fields

Currently there appears to be but one method of unstructured search (though there are many different variants of it) which involves the search for specific textual targets through textual contents.

Fast searches usually involve a structured database of significant words with pointers from each to their location in the document.

Structured searches rely on the data being appropriately organized or structured. We will use the term "File" to represent the element of information, though that could be any kind of information not necessarily a separate file and in structured databases it is usually the Data Record, but could also be the field of a record or some combination of fields taken from the same or even different records.

Information stored in computers is always structured to some extent. For example, all computer files are stored in hierarchical directories. This hierarchical structure is there to help organize the various files for easier access. The final step in user access, is a linear search through an alphabetized (or otherwise organized) list of names inside a directory. This is an example of a structured search. The ease with which a particular file can be found depends on how well organized the files are in various directories, how many directories there are, how diverse the data is and many other factors.

When organizing a system of directories the goal is to name them and the files descriptively so the user can find a file by following a descriptive path. When the number of Files is large, this goal is difficult to approach and as the number of files increases becomes impossible to achieve. The difficulty arises because the descriptions of files are not naturally hierarchical, and in addition, my hierarchy is not your hierarchy.

An example should help to illustrate this point. Imagine a set of emails, letters, and faxes received and sent by a user, stored in separate files organized into directories. There is mail from and to numerous countries of the world, many companies, universities, various individuals, and about different topics. The broad description just provided is usually the basis of the hierarchical directory structure of a well-organized file database. However, there is no unique prescription for the precise structuring of the hierarchy of directories. No such unique prescription is possible. Different arrangements of the hierarchy are equally convenient or inconvenient. For example:

*France:IBM:Legal:Letters:*

or

*France:IBM:Letters:Legal:*

or

*Letters:Legal France:IBM*

or any other ordering of the directories.

Not only is the structure of the hierarchy not unique, but the location of a particular file is often not unique. For example, a file which represents letters involving more than one topic may properly belong in two different directories, one for each topic. This is an example of the general overlap of topics which is common in files. Such overlap becomes greater the larger the database. The overlap problem is very familiar to anyone who has tried to organize a filing cabinet.

**Structured Data.** The traditional record-field databases provide another example of a structured database. If used to store files, the traditional way to retrieve information is through searches of specific textual content or specific field values. This is similar to the search through unstructured Files, in that text has to be typed and the search proceeds by searching the whole database for a match. The results are as usual hit-or-miss.

Searches through such databases can be made faster by creating a Table for each field to be searched. Often searches are needed on many different fields, and so many different Tables are required. The problem then occurs when the data structure needs to be changed in some way, as for example when an additional field or a connected record is needed. Usually this means a rather time-expensive process of re-indexing each of the Tables.

Compared with this, the TIE system runs on one universal Table which, because of its theoretical structure, will be referred to as the Matrix. It stores all the possible associations between Items in a very compact and access-efficient structure. Therefore the most time-expensive change that could possibly take place could require the re-indexing of just this one universal table.

It is important to realize that the changes introduced by the TIE system when implemented in structured databases, require a completely different approach to the subject - a completely different way of thinking about structured data then the present very inflexible state-of-the-art. Persisting in the old ways of thinking can lead to a misunderstanding of the possibilities and a likely non-optimal implementations of the technology.

**Unstructured Data.** Unstructured searches rely on the contents of the file. This has the problem that the user must know reasonably accurately some unique file content. Furthermore, no guidance is presented to the user as to what information is available. Typically the search is first defined by the user and then proceeds through the whole database often resulting in either nothing found or in too many found files. The principal problem with such searches is their hit-or-miss nature and the relatively long time they take. In large databases, it is often difficult to hit a search pattern which is sufficiently specific to be in only a small number of files but not too specific to miss what you are looking for or to find no files at all. There is also the additional difficulty of determining the best terms to search for and the inability to know if the information being searched is in fact present in the database.

TIE on the other hand, allows the user to describe the target file using a natural language description while being guided by immediate feedback.

Here is an example of a file description: **Received Email**, from **France**, about a **Distribution Contract** for a **Word Processor** called **Nisus.** A good search system would accept such a description and be able to almost instantly produce a list of file names which match it or instantly tell the user there are none. But even such a good system would have hit-or-miss aspects, because the user would need to go to the trouble of making up the description, without knowing if it is too narrow or uses the wrong words for there to be any match at all. Far better if the user is guided in some way while making up the description so that a hit is assured and no time is wasted looking for nonexistent information, or using the wrong terminology. Additionally, the guidance can show the user the available information.

TIE, as will be shown presently, is in fact such a system.

Referring to the example above suppose we describe the file using the key words (in bold) from the full description.  We get:

**Email, Received, France, Contract, Distribution, Word Processor, Nisus**

This is a list of descriptive key words and phrases which we call "Categories."

Great simplification in use results if we drop any significance to the order of such Categories in the description, something impractical to try in a hierarchically structured implementation.

This is the kernel of the simple idea behind TIE. In order to make sure that the user describes the file using only available words, a pre-defined set of Categories is presented in lists which can be organized either alphabetically, into logical groups, or in any other convenient intuitive way. These Categories can be thought of as the restricted vocabulary in terms of which the user is required to make up a description of the files being sought.

Avoiding the hit-or-miss process, the user makes up the description, progressively adding one Category at a time, while the Category lists are updated after each choice in the following way. When a Category is chosen, the remaining vocabulary of Categories, is immediately shortened by removing or disabling or in some way distinguishing, the unavailable Categories, that is those Categories which, if combined with the already chosen ones, would find no Files. In this way the user is progressively guided, in a non-hierarchical way, to the available information, avoiding any zero hit possibility.

Some may recognize that the system just described involves Boolean queries, using the conjunctive "AND" between the Categories. A more general implementation of this technology includes all possible Booleans involving all possible operators. A very convenient and practical approach to this generalization, described in detail elsewhere (in a concurrently prepared paper on the software implementation of the technology) is to present Categories in groups, where the kind of operator used implicitly with members of the group, is determined by the nature of that group.

The Boolean query picture of TIE is of course only less than half the story. The rest involves the determination of the available Categories as a result of the Boolean query. This part is by far the most processing intensive one and no present query system uses it.

This briefly describes the simple principles of the TIE technology. Surprisingly such a simple idea has extremely disruptive and profound consequences.

One of the important and powerful properties of describing files (really information objects) in terms of Categories is how few such Categories are needed to make the description totally unique, even in very large databases. This property is used in uniquely identifying each field and subfield of a record in a TIE implementation of new structured databases.

The use of a limited vocabulary of overlapping Categories to describe search criteria appears to be only a very slight and subtle change from normal hierarchical practice. Most casual observers of demonstrations of the technology are fooled into thinking that it contains nothing new. In fact, of course, it represents a major paradigm shift and has very profound consequences on both the analysis of data and the ease of information retrieval. The fact that it is a <u>limited</u> relatively fixed ("pre-defined") vocabulary of Categories, assigned appropriately, means that Categories can be used in an analytical framework of the system.

This paper begins by setting up such an analytical framework within which the consequences of assigning Categories to Items and retrieval criteria can be investigated. The last section on applications describes very briefly the very numerous applications of this technology.

## 2    The Association Matrix

The system we are investigating consists of a set of Items of information, which could be files, but generally referred to as Items, each described using a vocabulary of Categories. In what follows, we will use the word "File" and the word "Item" interchangeably.

In all implementations, it is convenient to use an ID number to identify each Item and an ID number to identify each Category. There is then a table connecting each Item ID number to the Item's location and/or its name, and another table connecting each Category name to its ID number. In many practical implementations Group Ids are also used, but they do not need to be taken into account in our discussion here, because they only concern the GUI.

Groups are used to logically organize the Category Vocabulary. Such a logical organization allows user selections of Categories to be logically and uniquely interpreted as complex Booleans involving all the usual Boolean operators, without the user ever needing to know about Booleans or the associated operators. The details of this will be presented in a subsequent paper on implementations and user interfaces.

The assignment of Categories to Items will be referred to as Categorization. How this assignment is done is the subject of non-hierarchical taxonomy (which is much easier to automate in software then the conventional hierarchical kind) and will be discussed in a subsequent publication.

Let N be the total number of Files in the system and n the total number of Categories. In most practical systems, the number of Files will be orders of magnitude greater than the number of Categories (N>>n).

Visualize the system of files with assigned Categories as a binary n by N matrix M in which the column labels are the identifiers of the Files and the row labels are the identifiers of the Categories[2]. An element of the matrix M is either 0 or 1. Each element expresses the association, or lack thereof, between a Category and a File, the value 1 signifying an association and zero signifying none. (In a more general system each element would be a number, representing a measure of the relevance of the Category in describing the associated Item.)

We will refer to this matrix as the Association Matrix or simply M.

We are free to arrange the ordering of the rows and columns of M in any convenient way. Having made some choice of ordering, we can change it by re-ordering the columns and rows without affecting anything physical, as long as during any rearrangement we keep the associations between Categories and Files unchanged, that is, we appropriately re-order the row and column labels, or update the name tables.

**The Ordered Matrix.** The following defines the *Ordered Matrix* M.

[2] I am greateful to Wojciech Fiedurek, a visiting engineer from Krakow, Poland in 1996 for suggesting this visualization.

Imagine a matrix M for a large file system. Typical numbers of files might be from several hundred thousand to 10 million or even billions, controlled by several hundred to several thousand or even hundreds of thousands of Categories. (It will be shown how extremely large numbers of Categories can be handled). The picture is that of a sparse population of ones, with mostly zeros in the n by N matrix M.

Total up all the columns and all the rows. Then re-order the matrix by moving whole columns so as to end up with the columns with highest totals on the left, with their totals decreasing from left to right, the last columns having the smallest totals. Now do the same for the rows rearranging them, so that the highest row totals are at the bottom, decreasing to the lowest at the top.

The sums of the rows represent the number of Files assigned each corresponding Category, whereas the sums of the columns represent the number of Categories assigned each corresponding File.

Think of each row of zeros and ones as a Category Vector, and each column as an Item Vector.

There may be some Category vectors which have identical components. Any such Category vectors will come only from subsets of Categories with the same row sum. Similarly for File vectors. We define the *Ordered Matrix* as that which, within subgroups of equal sum (in both columns and rows) brings equal vectors (that is, those having identical components) together first, followed by those with two components different, then those with four, and so on. (It s of course impossible to have binary vectors with equal sums differ by an odd number of components.)

It is useful to define various subsets of the row and column vectors. The subsets of those Category (row) vectors which have equal sums we call Equisum sets. Vectors of the subset of the Equisum set which have identical components we call the Synonym set when they refer to the row (Category) vectors, and the Inference set when they refer to the column (File) vectors. The Synonym set of Categories can be replaced by one category with the others being its synonyms. The Files within an Inference set share a common subject and in that sense infer each other. In some applications, it may be desirable to merge them into one file.

In this form the matrix will be referred to as *Ordered*. Visualizing it in this Ordered form, the population of ones is highest in the bottom left corner, the origin, when visualized as an x-y plot, decreasing to the lowest in the top right corner. $M_{ij}$ is then the element of this matrix giving the association between the $i^{th}$ Category and the $j^{th}$ Item.

**Graph Visualization.** This binary matrix can be visualized as a graph of connections between two types of nodes: the Items and the Categories. In the following example graph we use an unfilled circle to represent the Item and a filled one to represent the Category.

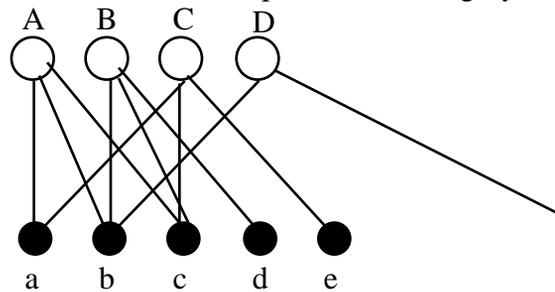

Fig. 1

Figure 1 graph represents the following assignment of Categories to Items:

A(a,b,c); B(b,c,d); C(a,c,e); D(b,...)

where the capitals stand for names of Items and lowercase letters for names of Categories.

One fundamental measure of M is the total number of ones (or connections) in the matrix, or the sum, S, of all the matrix elements:

$$S = \sum_{i=1}^{n} \sum_{j=1}^{N} M_{i,j} \qquad\qquad (1)$$

The matrix M or its programmed equivalent, is usually kept in RAM for quick access. This data, which will be referred to as *Matrix Data* is proportional to S and so the memory required to store it is also proportional to S, although in software implementations it is not always stored as a matrix but sometimes as a set of Category vectors, the components being the ID numbers of the Items. (This method of storage uses less memory and often performs better when the ratio of average number of Categories per Item to n, the total number of Categories, is less than 1/32 for 32 bit implementations.)

The number of Categories assigned to the $j^{th}$ File is $C_j$, the column sum, given by:

$$C_j = \sum_{i=1}^{n} M_{i,j} \quad (2)$$

The number of Files using the $i^{th}$ Category, the row sum, is $F_i$ given by:

$$F_i = \sum_{j=1}^{N} M_{i,j} \quad (3)$$

The average number of Categories per File is $C_{av}$, given by:

$$C_{av} = \frac{1}{N} \sum_{j=1}^{N} C_j = \frac{S}{N} \qquad\qquad (4)$$

The average number of Files per Category is $F_{av}$, given by:

$$F_{av} = \frac{1}{n} \sum_{i=1}^{n} F_i = \frac{S}{n} \qquad\qquad (5)$$

Clearly:

$$\frac{C_{av}}{F_{av}} = \frac{n}{N}, \quad (6)$$

which is a very useful, fundamental relationship.

In dealing with binary matrices, we define two kinds of addition: the normal decimal addition, which generally results in a non-binary result, and the binary addition which always results in a binary result. So that, for example, under ordinary addition $1 + 1 = 2$, under binary addition $1 +_B 1 = 1$. Binary addition is equivalent to the boolean "OR."

# 3    Categorization Quality Measures

$c_i^j = M_{i,j}$ can be regarded as the $i^{th}$ component of the n-component binary File vector $\mathbf{c^j}$ for File j where each non-zero component $M_{i,j}$ means that the $i^{th}$ Category is assigned to the $j^{th}$ File. A different, non-binary version of this File vector is almost always more practical to implement in software. It is a vector with components which are the IDs of the Categories associated with the given File.

Alternatively $c_i^j$ can be regarded as the $j^{th}$ component of the N-component binary, Category vector $\mathbf{c_i}$ for Category i. There is likewise a corresponding non-binary Category vector, whose components are the IDs of those Files assigned the given Category. These File IDs are always sorted in increasing order so that boolean comparisons can be carried out very quickly.

We will use the binary product of vectors, defined by $c_i = a_i b_i$, written as $\mathbf{c} = \mathbf{ab}$, that is, the boolean "and" of corresponding components when the vectors are both binary vectors.

The sum of $\mathbf{c^j}$, defined by $Sm(\mathbf{c^j}) = \sum_{i=1}^{n} c_i^{\,j} = C_j$, ............................................................. (7)

is the number of Categories assigned to the File j. Similarly for the vector $\mathbf{c_i}$,

$$Sm(\mathbf{c_i}) = \sum_{j=1}^{N} c_i^{\,j} = F_i, \hspace{3cm} (8)$$

which is the number of Files assigned the Category i.

We now setup two useful parameter measures (q and r) of Categorization quality.

Let $q_j$ be the number of Files <u>having exactly the same set of Categories as File j, and no others</u>. Call this the *Tight Inference Number* because it counts the number of files that infer each other regarding their overlapping subject matter to the maximum degree allowed by the pre-defined set of Categories: it is the number of Files in that Inference Set which contains the file j. This is the *tight* inferencing definition - in the sense that the inferenced files have completely overlapping descriptions within the given Categorization system. (The most loose definition would consider any files with just one Category in common to infer each other.)

The algorithm for calculating $q_j$ can be deduced as follows. First sort the Matrix elements into an Ordered Matrix as defined above, then count the number of elements in each Inference set to get each $q_j$. The average and standard deviation from the average of $q_j$ as well as the set with a maximum $q_j$, can then be used as measures of the Categorization quality. Perhaps the most useful of these measures is the maximum $q_j$ call it $Q_j$ and the associated file set.

Categories are intended to distinguish the Files so that it is easy to locate any particular file. To meet this criterion, it is imperative that this $Q_j$ be no more than say 10 or 20, and the smaller it is the better. It represents the number of Files listed when the relevant Categories are selected by the user. Locating the File being sought is then easy by its name in the listing - the name used as the final, detailed defacto distinguishing Category. If the list of files is too long, locating the target file will not be as convenient.

Files with a $Q_j$ greater then the desired low number can be Categorized as "Badly Categorized" ready for manual editing of their categorization.

The File to Category transpose, $r_i$ is defined similarly. $r_i$ represents the number of Categories that are synonymous with Category i. For greatest efficiency this number should be 1 for every Category. That is, no Categories should be synonymous with any given Category. However, in practice it may be convenient in some applications to use such synonyms.

In practice, another parameter, *Granularity*, $G_j$, may also be a useful measure of quality of categorization. Granularity, defined for each file, is also the number of files which have the same categories as the given file, but may also have additional categories. For practical reasons, in well categorized systems, like $Q_i$, $G_j$ too needs to be smaller than about 20 for any file. The files for which this number is excessive can be automatically assigned some descriptor Category like "Needs Better Categorization" enabling them to be easily found and their categorization edited.

A useful visualization parameter is the density of non-zero elements in the M-matrix. This density, $a$ is given by $a = S/(nN) = C_{AV}/n$.

In a typical, well-categorized database using about 700 Categories, and typically having a $C_{av} = 7$, $a$ is about 0.01, which means that 99 times as many elements in the M-matrix are zero as are non-zero.

# 4    Squaring the Matrix

The square of the matrix M (which can be thought of as representing correlations between the matrix elements) gives rise to interesting physical quantities. For example, in the n by n File matrix has elements

$$F_{i,i'} = \sum_{j=1}^{N} M_{i,j} M_{i',j} \qquad (9)$$

Each element represents the number of Files which have assigned them both Categories i and i'.

Similarly, in the corresponding N by N Category matrix

$$C_{j,k} = \sum_{i=1}^{n} M_{i,j} M_{i,k} \qquad\qquad (10)$$

each element represents the number of Categories which are assigned to both Files j and k.

These matrices are no longer binary, but they do have their binary counterparts. The binary version of F has elements which if non-zero, indicate that the corresponding Categories have at least one file in common, and if zero, indicate no files in common. Similarly for the binary version of C and Files having Categories in common.

A very useful application of the binary version of F is in caching the available Categories for all the initial clicks. I turns out that in most very large systems the first click - that is, the choice of the first Category - requires the most processing time. So it is useful and quite economical to cache all first clicks. The F matrix is the information which needs to be cached.

The diagonal elements of both these (non-binary) matrices are the same as the expressions linear in M, so that $F_{i,i} = F_i$, is the File or row sum, representing the number of Files assigned the Category i, while $C_{j,j} = C_j$, is the Category or column sum, representing the number of Categories assigned to the File j.

Consider the (sample) averages of the File and Category matrix elements, obtained by summing over both indexes and dividing by the number of terms in the sum. The Category matrix element average is the average number of Categories assigned to any pair of files. It can be shown to be related to the standard deviation, $\sigma_F$ of $F_i$ as follows.

$$< C_{j,k} > = < \sum_{i=1}^{n} M_{i,j} M_{i,k} > = \frac{1}{N(N-1)} \sum_{j=1}^{N} \sum_{k=1, k \neq j}^{N} \sum_{i=1}^{n} M_{i,j} M_{i,k} = (n<F^2> - S)/(N(N-1)) \dots\dots\dots (11)$$

and this can be expressed in terms of $\sigma_F$ by using the result

$$<F^2> = \sigma_F^2 + F_{AV}^2 \qquad\qquad (12)$$

so that

$$<C_{j,k}> = n \; \sigma_F^2/(N(N-1)) \qquad\qquad (13)$$

Similarly, evaluating the corresponding average of non-diagonal elements of the Files matrix (or simply using the Categories <---> Files transpose) gives the result:

$$<F_{j,k}> = N \; \sigma_C{}^2/(n(n-1)). \qquad\qquad (14)$$

From these relations follows the ratio between the corresponding Category and File averages:

$$\frac{<C_{j,k}>}{<F_{i,i'}>} = \frac{n^3\sigma_F^2}{N^3\sigma_C^2}, \qquad\qquad (15)$$

where we have approximated using n in place of n-1 and N in place of N-1.

# 5    Narrowing Factors

**Category and File Narrowing Factors**. When designing a TIE system it is useful to make estimates of the rate of narrowing of the file hits after each Category selection and the corresponding rate of narrowing of the remaining listed Categories.

We define the File Narrowing Factor of a Category as the ratio of the number of matching files before the Category was selected to that number after the Category is selected.

In practice when designing a TIE system we have control over the minimum and maximum number of Categories assigned to any file. Through this, we control the average number of Categories per file, $C_{AV}$, which must lie between the two limits. Indirectly, through $C_{AV}$, because of equation (6), we control the average Files per Category or $F_{AV}$, which is the more directly relevant number in calculations.

In the average situation, the selection of the first category will result in the total number of Files N being reduced to the average number of Files per Category or $C_{av}/n$ times N, using relation (6). So the reduction factor, or File Narrowing Factor, for the average case is $C_{av}/n$.

Before estimating the Narrowing Factor when the second Category is selected, consider the estimate of the Category Narrowing Factor. In a similar way to the File Narrowing Factor, we define the Category Narrowing Factor associated with a given Category, as the ratio of the number of listed Categories before the selection of the given Category, to that number after its selection.

To estimate it on the assumption of a completely random Category to File assignment, we first note that the probability ($p_{ii'}$) of any given Category identified by the index "i" of having at least one file in common with another Category identified by the index "i'" is given by:

$$f_{ii'} = \frac{F_i F_{i'}}{N^2} \qquad (16)$$

because $F_i$ is the number of files associated with Category i and N is the total number of files.

Therefore the probability of Categories ($f_i$) which have at least one File in common with the Category i is given by:

$$f_i = \sum_{i'=1, i' \neq i}^{n} f_{ii'} = \frac{SF_i}{N^2} - \frac{F_i^2}{N^2} \qquad (17)$$

where $F_i$ is given by equation (3). Averaging both sides over all Categories, using equations (4) and (5) gives:

$$f_{av} = \frac{C_{AV}^2}{n^2}(1 - \frac{1}{n^2}) - \frac{\sigma_F^2}{N^2} \cong \frac{C_{AV}^2}{n^2} \qquad (18)$$

the approximation being equivalent to neglecting the diagonal terms in the summation. It is justified because $n \gg 1$ and $\sigma_F \ll N$.

Following the first Category choice the narrowing factor was $C_{AV}/n$. Therefore the Categories Narrowing Factor when the second Category is chosen, will also be on average, $C_{av}/n$, which is exactly the same as Category Narrowing Factor following the first Category choice and the same as the Files Narrowing Factor.

This estimate assumes a completely random Category to File assignment, so in practice its predictions cannot be relied upon. In practice,, we have found that in systems with large numbers of Items and a uniformly (not randomly) assigned system of Categories to Items, it often happens that the First Category selection produces hardly any narrowing of Categories, and on the subsequent selections produce substantial narrowing. A simple example should make this more obvious.

Suppose our system Items represent events uniformly distributed through the days of the week. Suppose then that each item is described by a set of Categories, one of which is the day-of-the-week Category. If there is at least one event described by each kind of Category, occurring on each day of the week, a selection of a day of week category will not narrow the remaining category vocabulary - though of course it will narrow the Item list.

Obviously the smaller the Narrowing Factor the smaller the number of selections before you find what you are looking for and the better the Categorization system. Therefore it is important to design Categorizing systems which have a sufficiently low $C_{av}/n$.

Consider the possible range of $C_{av}$. Assuming that all files are assigned a Category, this number must be greater than 1. Assuming that the assignment is reasonable and useful that number must be less than n. In most practical systems, judging from our experience with several large systems we expect this average to be between 4 and 50 in systems where the data is taken form a structured database. In unstructured systems where the Categorization is automatic and no limit is placed on the number of Categories, the average could be as high as several thousand.

The total number of possible Category combinations is $2^n$, which is the total possible number of files that could be uniquely Categorized using n Categories. Just a 100 Categories can Categorize $10^{30}$ files! However the actual number that can be usefully Categorized using a particular set of n Categories, is usually a very tiny fraction of that.

(A more realistic, though still not very practical, estimate would estimate the number of possible ways to pick 10 Categories out of 100, giving about $10^{13}$, still a very large number.)

Reducing the Narrowing Factor $C_{av}/n$ means reducing $C_{av}$ and increasing n. Reducing $C_{av}$ by simply assigning fewer Categories to each File, is not practical. In fact as n increases, $C_{av}$ must also increase if the added Categories are useful and if they are not needed to describe new Item topics.

Assuming each of the n Categories is descriptive of some of the Files, when assigning Categories to Files we are obliged to assign every relevant Category, otherwise some resulting searches will be futile. So effectively, we have no choice in limiting the Category assignment. The only way, therefore, to reduce $C_{av}/n$ is to increase n while limiting the possibly resulting increase in $C_{av}$.

Simply stated, this means using descriptive Categories which are just of the right breadth in the number of Files they describe. For if the descriptions are too broad, we will need too many Categories for each File to be able to discriminate between Files, hence increasing $C_{av}$. If the descriptions are too narrow, our Narrowing Factor will be very small and our discrimination will be very good, but the total number of Categories (for some applications of the technology) will be too large to be practical and in the limit could approach, or even exceed the total number of Files, N.

Automating the optimization of the choice of the vocabulary of Categories is a very complex one, requiring a great deal of computing power, and may be impractical to completely automate except using genetic algorithms and programs. In fact this technology is probably ideally suited to such algorithms, because the various Categorization quality parameters can be used as tests and others can be added to check the quality of Categorization.

# 6    A simple Analytic Model

It is useful to have some means of modeling the system of Files and Categories and their assignments, so that we can perform calculations of the various parameters.

The following describes such a simple model.

We begin with the binary Ordered matrix M described above. We assume the number of files is sufficiently large so we can make meaningful estimates using statistical calculations.

Because we have arranged it so, the origin of the M matrix (bottom left corner, where both the Category and the File index start at 1) is the region of highest density of the dots (representing 1's) and the density of these dots decreases with distance from the origin.

Each column sum, $C_j$, the number of Categories assigned to the file j, decreases from a maximum ($C_1$) at j = 1 to a minimum ($C_N$) at j = N, without ever increasing. Similarly, the row sums, $F_i$, representing the number of Files with the Category i assigned to them, are by arrangement also a maximum ($F_1$) at i=1 and then decrease to the minimum ($F_N$) at i = n, never increasing.

Such behaviors can be modeled using some simple functions and realistic situations. We illustrate two simple models both of which give similar values of the Narrowing Factor.

In practice, when categorizing a large set of Files using a specified set of Categories, the person performing the Categorization, implicitly or explicitly, is instructed to assign no less then some minimum number of Categories ($C_N$ for an Ordered matrix M) and no more than some maximum number ($C_1$ for an Ordered matrix M) of Categories to each File.

Furthermore, the instructions include the statement that of the available Categories, all possible applicable Categories must be assigned to each File. Therefore, we can assume that if the Categorization is done properly, all Files are Categorized and no Categories are unused.

Our crude models approximate $C_j$ as a continuous function using polynomials. The simplest, linear function is given by:

$$C_j = C_1(N - j)/(N - 1) + C_N(j - 1)/(N - 1) \qquad (19)$$

With this linear model the relationships are as follows:

$$C_{av} = S/N = C_1[N/(N - 1) - 1/2] + C_N[1/2 - 1/(N - 1)], \text{ or assuming that N is very large the simple,}$$
obvious result

$$C_{av} = S/N = (C_1 + C_N)/2. \qquad (20)$$

Using a quadratic for $C_j$ with the same assumptions, gives the results:

$$C_j = C_1 - j^2(C_1 - C_N)/N^2 \qquad (21)$$

from which we get:

$$C_{av} = S/N = 2C_1/3 + C_N/3 \qquad (22)$$

where we have made the additional assumption that the derivative of $C_j$ is zero at the starting point $j = 1$ (which, because of large N, might as well be $j = 0$). This assumption is justified by the fact that, striving to meet the instructions, the categorizer will try to maximize the number of Categories assigned to each File, hence "bunching-up" the number of Categories at and near the $j = 1$ region which, because of ordering, is by definition the region with the highest values of $C_j$.

The mean square can also be estimated. Using the linear model, we get:

$$\langle C^2 \rangle = (C_1^2 + C_N^2 - C_1 C_N)/3 \qquad (23)$$

keeping only leading terms in N, for large N.

From this, the square of the standard deviation is given by:

$$\sigma_c^2 = <C^2> - C_{av}^2 = (C_1 - C_N)^2/12 \qquad (24)$$

Similar expressions are obtained for $<F^2>$ in terms of $F_1$ and $F_N$,

Interestingly, the quadratic model gives structurally and numerically very similar results, with

$$\sigma_c^2 = 4(C_1 - C_N)^2/45 \qquad (25)$$

Consider some examples. In a test Categorization of about 200,000 web pages using about 1,000 categories, experimentation resulted in instructions to the categorizer to hold the minimum number of assigned Categories to any file at 4 and the maximum at 10.

With these numbers, the linear model gives $C_{av} = 7$ and so the Narrowing Factor after each click on a Category, on average, would be 7/1,000. Therefore the first click would reduce the 200,000 web pages listed, on average, to 1,400, which on the second click would reduce, on average, to about 10! So, on average, in such a relatively large database, two clicks are sufficient to find any web page!

Using the quadratic model, the results are very similar, so that $F_{av} = 8$ and the Narrowing Factor is 8/1,000.

Clearly the worst possible case of an average narrowing factor is 1/100 in this example, because the corresponding largest $C_{av} = 10$, which gets the Web list down to 200 after two Category selections, on average requiring only a third Category selection.

All these estimates use average quantities so of course there will be many instances of more than two or three Categories needed to narrow down the File list sufficiently. In practical even badly over-categorized systems like the example above, it was found that very few cases required more than about 4 Category selections.

Consider the simplified analysis of the "dot" statistics. Because the maximum and minimum Categories per File are relatively close to each other, the density of dots in the Ordered matrix is relatively uniform, with only a small decrease from the origin.

# 7    Very Large Data Bases

TIE is the ideal solution for handling very large databases. Specifically the most difficult problem is that of hundreds of millions of data items each of which is an unstructured file, such as a word processor document or a presentation or any other document containing text, including a web page. We assume, therefore that we have, for example, one billion documents containing text and that we have to find a way to Categorize all these using suitable key words as Categories.

TIE implementations work best, with the fastest response, when the index matrix is held in and accessed entirely from memory. The memory requirement of the typical such index matrix, the Matrix Data, is proportional to S as given by equation (1) and can be estimated as $4*C_{AV}N$ bytes, where $C_{AV}$ (see equation (2)) is the average number of Categories assigned per information Item and N is the total number of information items. (This memory requirement is doubled when using the so called "Double Matrix" implementation for faster response times.)

For one billion files (information items) the number of Categories needed can be as little as 1,000 or as large as 300,000 (the typical English dictionary). It is important to remember that, the total number of Categories is not what determines the Matrix Data size when using the ID vector implementation of the matrix) - rather it is the average number of Categories per Item.

If the subject matter in the database spans a very broad subject domain, such as for example all the subjects in the Library of Congress, it will be difficult to limit the Categories to 1,000.

So our premise is that we will need practically the whole of the English language dictionary of words as our Categories. Although the unabridged dictionary is considerably larger than 300,000 words, most of those words would not be needed in a set of Categories. Therefore let us assume that we will need no more than about 300,000 Categories.

Taking the worst possible case of using all the significant words in each document as the Categories, we assume that the typical average document contains about 1000 unique, significant words. Therefore the storage needed for the index matrix would be $4*1000*1,000,000,000$ bytes or 4 terabytes. Currently such storage can be accommodated but only on hard media. To require this much RAM is currently unreasonable (though in the near future it will become common). If we assume a more moderate number of unique words per document, say 200, then the storage required is a more moderate 800 GB.

Therefore such an Index matrix should be created and stored on disks. The access to this can be very fast to return the found Items. However it is the return of the available Categories that takes most of the time, and this would probably be too slow. However, the most severe demands on time occur when the first Category is chosen. Such single Category choices could all be cached in a Category-Category matrix which would require much less memory, particularly when optimally stored. Such issues are discussed in a subsequent publication dealing with applications.

Another solution to the very large numbers problem is to use parallel processing. The TIE matrix can be very easily divided into parts, each one being serviced by a different processor and each one dealing with a subset of Items using, where appropriate the whole set or almost the whole set of Categories. The results from each are then easily combined (simple addition is all that is required with negligible processing demands).

Parallel processing can also be easily achieved by using a large number of independent servers with a client broadcasting to all servers each request, and combining the responses to present to the user. Therefore a billion Item database can be quite conveniently distributed amongst, say 1,000 independent PC servers, each serving only a mere 1 million Items.

**Textual Analysis**. Typical, what is needed when unstructured text documents are to be Categorized is the textual analysis of all the documents with the intent of extracting the relevant words as Categories, excluding a stoplist of irrelevant words. The words are analyzed for their frequency of use within each document and within the world of documents. Various relevance indexes are calculated.

As an example, the simplest relevance index would be the percentage of documents which use the given word at least once. When this percentage is above a certain threshold, the corresponding word is a good Broad Category. If this percentage is very low, the corresponding word becomes a good Detailed Category. Of course the transition between these two extremes is almost continuous and other relevance indexes can be used.

**Top Level Categories.** To make a system requiring a large number of Categories per Item and a large number of Items work without requiring an overwhelming amount of RAM for holding the matrix, we can initially present the user with a reduced number of Categories - the Top Level Categories (TLC) - and when the number of remaining Categories is sufficiently reduced present the rest as a final step.

The Top Level Categories can be selected using their Popularity index - that is the number of Items that are assigned the Category, or by the Relevance Index of Category to Category connections and/or by manual editing of the list.

The set of TLC can be arranged to be a small fraction of the total number of Categories as the example will illustrate.

The TLC portion of the total matrix will be designated as $M_D$ and referred to as the Dominant part of M.

The following objectives are required of the TLC:

1   The number of Categories in the TLC must be as small as possible, consistent with the remaining objectives;

2   The subset must be representative of all the topics in the database. Equivalently, it must be possible to describe any broad topic covered by the database by overlapping Categories from the subset.

3   It must be possible to narrow the full set of Categories to a sufficiently short list by selecting a reasonable number of the Categories from the subset.

A powerful and very logical method of selecting the TLC subset, can be generally described as follows:

Objective: The Subset must be Maximally Discriminating of the remaining Categories. This means that the subset is the smallest subset of Categories which will allow a suitable combination of them to access all remaining Categories in groups less then some number like 1000.

First use the criterion outlined here, that is using the best relevance index, to pick  several times the size (p) needed, ordered with the highest relevance score first. Then starting with, for example, the first 100 Categories of highest relevance, test to check if all remaining Categories can be "accessed" by using some combination of these and if each such combination leaves behind no more then say 1000 of the remaining Categories to pick from. If the picked subset does not achieve this, then check by adding one Category at a time from the remaining group, in order of relevance, and add those which improve the discrimination of the lacking subsets, until the objective is achieved. This needs some fast algorithms to make it practical, because it is an n! process. This is where a genetic algorithm might work very well.

The user is instructed to begin the search by choosing from the TLC subset. The $M_0$ matrix which (it will be shown) can be easily held in memory, allows for instant response, removing the irrelevant Categories and reducing considerably the remaining relevant DCs. Once the list of available Categories is sufficiently short, the DCs can be displayed for the user to complete the detailed search. At that point the system can check the master matrix from a disk file, determining the number of hits, and display that information. The user can then decide to continue narrowing through further Category selections or to make an immediate choice to view a data item.

The total number of Categories is limited to the total vocabulary. Therefore as more File items are added, the vocabulary grows more slowly, reaching asymptotically the total vocabulary. The average number of topics that overlap each other will usually not continue increasing, but will top out at some reasonable value.

The assumption of 1000 as the appropriate number of TLCs is arbitrary, but of the right order of magnitude. In essence these 1000 TLCs are sufficient to describe the general subjects contained in the database. So for example, a typical library will have no more than several thousand top level subjects.

**An Experiment.** In an experiment using about 840,000 newspaper articles as the data Items, it was found that choosing Capitalized words, not at the beginning of sentences, excluding the words on a short stoplist, produced a list of about 350,000 words. The analysis of the text calculated a new relevance factor for each of these words by counting the cumulative number of links with other words in the set introduced by each article. So that each time a given word was found with other relevant words the relevance count for that given word was incremented by the total number of relevant words in that document less one. This turned out to be a good measure of relevance. It is intuitively the number of connections to other words that each document establishes.

We then picked the 3000 words with the highest relevance score, as the Broad Categories or the Top Level Categories (TLCs). The remainder we call the Detailed Categories or DCs.

For most applications the TLCs were sufficient to find any article on any topic. For the case when this does not suffice, the user would be presented first with the TLC to pick from and only present the DCs after the remaining DCs list is reduced to a manageable number through the narrowing process as Categories are selected from the TLCs.

Alternatively, we can divide the Categories into three or four "levels" the top level being those with the highest relevance index, with the index decreasing through the levels.

# 8    Other Applications of TIE

The essence of the TIE engine lies in the following essential fundamentals:

1    The mapping of a set of Categories to an information Item;

2    The representation of this mapping through a compact, optimized form of the matrix;

3    The calculation, sufficiently quickly, of the available Categories in each of any number of different type groups, using this matrix.

The fundamental property that only a small number of Categories is needed to make this mapping select or Filter a single Item (though in practice the less stringent selection of a few items is sufficient) from a very large number of Items, makes this system very powerful and applicable to many diverse fields. The retrieval of information from a large database is the one focussed on in this paper. In this section we briefly describe other applications.

**General**. This ability to Filter quickly and intuitively using the Matrix can be used in every other case where a large number of Items of information needs to be organized for access. There are many obvious fields where this is so and there are those not so obvious.

**The Brain.** Mapping the human brain using the Matrix is a less obvious application. It is well evidenced that human brain functions are localized and associated with overlapping Categories (See note [, page 3).The TIE matrix could be the ideal structure to use in mapping out all the complexities of brain function as they are discovered. Once carried out,this process would lead to certain testable consequences which could verify the system or require its modification.

**Pattern Recognition.** Pattern recognition is another less obvious application of TIE. Suppose we are to develop a system for recognizing a pattern out of a large selection of possible patterns. An obvious example is that of character recognition, which we shall use as an illustration.

The decision to use TIE immediately focuses us on the crucial "nuts to crack" to solve the problem: we must identify a set of sub-patterns within each pattern, sufficient in number and kind to uniquely identify each character. (In addition we may decide to also use attributes of the pattern placement as categories.)

Here we immediately see the pervasive and self-recursive nature of TIE system applications. In the process of identifying a sufficient set of sub-patterns, we need some form of cataloging tool. This tool must be able to tell us unambiguously if the newly identified sub-pattern, when added to the existing sub-pattern set, will help distinguish the various patterns from each other or not. It must further guide us in the search for those sub-patterns.

Not surprisingly such a software tool can be easily built using the TIE system. In fact a software application we developed many years ago called MailKeeper and meant for capturing, storing of email messages as well as any textual information, for easy retrieval, can be used without any change for this and most other cataloging chores. Here is how it would work in this context.

Each character would be an Item of information. As a sub-pattern is identified, it is named using any convenient word or phrase (in an application created specifically for the purpose, a graphic could be used and would be more appropriate). This becomes the Category. Each character containing a sub-pattern would be assigned that sub-pattern as a Category in an overlapping way as is usual in TIE systems. At any time in the process, we could simply click on an Item (Option-Click in the MailKeeper application) to be immediately shown all the sub-patterns associated with it and any other Items which have the same set of Categories. The objective is then to find more unique sub-patterns to distinguish those Items from each other. The process of pattern identification would be simulated by clicking on the Categories and noting how the list of possible Items shrinks after each click.

A good character recognition system can be built up by arranging for several redundant extra sub-pattern Categories associated with each Item, so that any sub-set of each Item's Categories is all that is needed to uniquely identify it. A Redundancy factor can be defined as a measure of this: a possible definition is the ratio of the total number of Categories assigned to the Item to the minimum size of the subset of these needed to identify the Item. A special application can be built using the TIE system which is more appropriate then MailKeeper, having more specialized features.

Once the best set of Categories is built-up, the software for the pattern recognition system would use the TIE Matrix in the process of identifying each character or Item. Furthermore, testing the system would be greatly facilitated by displaying each Category sub-pattern as it is identified, immediately also displaying the narrowing list of possibilities of the filtered patterns. Such dynamic display would help identify any problems of character discrimination and the reasons behind them.

A way can now be seen, using the TIE system, to develop software that will automate the process of identifying the best sub-patterns for best identification of the patterns.

**Sound Recognition**. When sound is converted to a pattern, sound recognition can be treated the same way.

**DNA Cataloging.** The identification of DNA chains is in many ways just another form of pattern recognition, so very similar methods can be used and very similar software tools can be built to help in the process.

There can be many different "levels" of Categories when dealing with DNA. At the lowest level, each letter in the set AGCT would be a separate Category, then each sub-sequence of interest when encountered could be added as a new Category. Automatic Categorization would be through literal matches - much easier then the text applications of TIE. The information Items are the DNA sequences. You click on the Categories (or sub-sequences) that you are searching for and all Items which contain these are immediately listed.

**Access to Long Lists.** In many databases you may need to select a name or serial number from a long list. In the case when the list of Categories is long, typing is the preferred method of access. In all the cases TIE system can be used to provide real-time immediate feedback of the selection as the user types each character.

The system works by using each individual letter of the alphabet (usually without regard to case) as a Category. After every keystroke the list of names, words, or ID numbers shrinks to show only those that contain the characters typed. When the user indicates the completion of typing, the remaining untyped character Categories are automatically set to be negated, which narrows the list of possibilities further.

This method of word/phrase selection has great advantages over current methods which need to perform standard searches. First it is very fast, second it is very lenient on common typing and spelling errors. The most common typing or spelling errors involve the transpose of letters and the duplication of letters. In the TIE system, each character is a Category, and as the order of Category selection is not significant and second selections of the same Category are ignored, the input will automatically correct for both such error types.

A simple test of such a system was performed on a list of about 60,000 personal names. Name selection was very fast and very lenient. In most instances after typing about 5 characters the list would shrink to a very small number of possibilities.

In such applications, other features that could help the ease of selection might include the indication, through a display in distinctive style or selection, of the exact match from amongst all the possible matches. Once the list is sufficiently narrowed, the search for an exact match would be very fast.

**Structured Databases.** Current databases are very inflexible and require the creation of many tables for fast searches on different field parameters. Developing a modern database system by applying the TIE principles would easily overcome both these problems. In addition, the TIE system dictates a very obvious and intuitive user interface.

Using the TIE system, each Record is Categorized with a set of descriptive overlapping Categories. In addition a unique Identifier, such as an ID number, can be used to uniquely distinguish each Record.

In general, three types of Category can be defined. Two of them are the Field Description Category (FDC) and the Field Value Category (FVC).

The user can decide which field values are to be used as FVCs and which descriptive terms should be used as FDCs. Categories can then be grouped and new Categories describing each group can be introduced as Group Description Categories (GDC).

For example, suppose a customer database is developed, then typically, the data would include Customer Information, consisting of data such as first and last names, billing and shipping addresses, email addresses, city and zip code. Other data would include information about Customer Preferences, such as products purchased and other preferences.

In this example, the words:

*Customer*

*Information*

can be defined as the GDCs either separately or together as a phrase. The words:

*Name*

*First*

*Last*

*Address*

*Email*

*........*

could be used as the FDCs. This list of Categories should contain a sufficient number of words to describe all the existing fields uniquely, so that the description of a field, using the overlap of these Categories, is the entity which uniquely identifies the field. When a new field is added, if the existing Categories do not describe it uniquely, new Categories can be added to make its description unique.

The restriction of current databases, that all records must have the same field structure, is totally unnecessary. Any record can have a new field added, without the need to change anything in the other records. The structure of each record can be different, and the relationships of fields within records is tracked entirely through the Categories.

Categories can also automatically track the relationships between different databases, so the relational aspect of databases is automatic and easily extended. Such tracking through Categories is also much more intuitive to the user because all they have to do is describe by clicking on a vocabulary, any attributes they are interested in and any existing relationship will immediately be apparent.

**Pen Input Methods.** The use of palm size computers requires new devices for inputting text. One of these, the Graffiti method, is based on a so called Unistroke method. Each character is input by the user using a single stroke of the pen. This requires some learning. Most of the characters are easy to learn and remember,but some are a little harder. So an alternative method would allow the user to print each character in its capital form. This might be termed the multi-stroke method. Now it is a matter of pattern recognition, where each printed character is split into appropriate sub-patterns, each of which acts as a Category. Only a small number of stroke-Categories is needed to distinguish each character of a large alphabet.

The useful interactive implementation of the pen input method using TIE would display the alphabet above the pen input window and as the user draws each stroke would narrow the display to those characters that contain those strokes. The user would be able to know if the interpretation was correct or not, and could complete the selection of the character with a quick tap of the pen on the appropriate character in the reduced list.

**Computer Filing Systems.** All present operating systems store their files in hierarchically directories.

This was adequate when the number of files was not too large and the topics covered did not appreciably overlap. Present day computer disks store an extremely large number of files.

The TIE system can be used to develop a very much better, non-hierarchical model for file storage. Storing a file would no longer need a folder within a folder path but would simply ask for a description, in terms of a pre-defined, though editable, dictionary of terms. Finding any file would be similarly simple - just click on the terms which describe it.

# 9    Conclusion

The flexibility and power of the computer has enabled us to bring together an enormous amount of information. This has greatly complexified information systems of all kinds. In the initial stages of this evolution, hierarchical systems were used to manage the information, but as the information increased in breadth and size, the hierarchy broke down and became inadequate.

We are not born to think hierarchically - we think in associations, where the associations have a tremendously complex overlapping structure. A hierarchical system cannot cope with this, and a new system is needed. We believe that the TIE system is it.

The TIE system introduces a subtly new way of programming associations between information items. This leads to their easier and more intuitive accessibility. It has very broad applications across all information systems and could begin a new revolution in programing both structured and unstructured database systems.